\documentclass[11pt,twoside,onecolumn]{article}
\usepackage[]{latexsym}
\usepackage{epsfig}
\usepackage{amsmath,amssymb}
\newcommand{\be}{\begin{eqnarray}}
\newcommand{\ee}{\end{eqnarray}}

\renewcommand{\d}{\mbox{${\rm d}$}} %d differenziale non corsivo in math mode
\newcommand{\lp}{\ell_{\rm p}}
\newcommand{\mpl}{m_{\rm p}}
\title{\bf Minimum black hole mass from colliding Gaussian packets}
\author{
R.~Casadio$^{a,b}$\thanks{casadio@bo.infn.it},
$\ $
O.~Micu$^{c}$\thanks{micu.octavian@gmail.com}
$\ $
and
A.~Orlandi$^{a,b}$\thanks{alessio.j.orlandi@gmail.com}
\\
\null
\\
$^a${\em Dipartimento di Fisica, Universit\`a di Bologna}
\\
{\em via Irnerio~46, 40126 Bologna, Italy}
\\
\\
$^b${\em Istituto Nazionale di Fisica Nucleare, Sezione di Bologna}
\\
{\em via Irnerio~46, 40126 Bologna, Italy}
\\
\\
$^c${\em Institute of Space Science}
\\
{\em P.O.~Box~MG-23, Ro~077125 Bucharest-Magurele, Romania}
}
\begin{document}
\maketitle
\begin{abstract}
We study the formation of a black hole in the
collision of two Gaussian packets.
Rather than following their dynamical evolution in details,
we assume a horizon forms when the mass function for
the two packets becomes larger than half the flat areal radius,
as it would occur in a spherically symmetric geometry.
This simple approximation allows us to determine the existence of
a minimum black hole mass solely related to the width of the packets.
We then comment on the possible physical implications,
both in classical and quantum physics,
and models with extra spatial dimensions.
\end{abstract}
\setcounter{page}{1}
\section{Introduction}
\setcounter{equation}{0}
The subject of gravitational collapse and black hole formation in classical
general relativity dates back to the seminal papers of Oppenheimer and
co-workers~\cite{OS}.
Since then, the literature has grown immensely, but the topic remains remarkably
complex (for an overview, see, e.g., Refs.~\cite{joshi,Bekenstein:2004eh},
and references therein).
\par
One thing we can safely claim is that gravity will come into play strongly whenever
two localized matter states approach each other to a sufficiently short distance.
Indeed, in 1972, K.~Thorne proposed the so-called
{\em hoop conjecture\/}~\cite{Thorne:1972ji}:
\begin{quotation}
\noindent
A black hole forms whenever the impact parameter $b$ of two colliding objects (of
negligible spatial extension) is shorter than the radius of the would-be-horizon
(roughly, the Schwarzschild radius, if angular momentum can be neglected)
corresponding to the total energy $M$ of the system, that is for~\footnote{We
shall use units with $c=\hbar=1$,
and always display the Newton constant $G=\lp/\mpl$, where $\lp$ and $\mpl$
are the Planck length and mass, respectively.}
\be
b\lesssim \frac{2\,\lp\,M}{\mpl}
\ .
\label{hoop}
\ee
\end{quotation}
The conjecture has been checked and verified in a variety of situations.
Of course, it was initially formulated for black holes of (at least) astrophysical
size~\cite{payne}, for which the very concept of a classical background
metric and related horizon structure should be reasonably safe
(for a review of some problems related to the hoop conjecture,
see the bibliography in Ref.~\cite{Senovilla:2007dw}).
\par
Whether the hoop conjecture can also be trusted for masses approaching
the Planck size, however, becomes more questionable.
In fact, for such (relatively) small masses, quantum effects may not be
neglected (for a recent discussion, see, e.g., Ref.~\cite{acmo})
and it is conceivable that (semi-)classical black holes must be replaced by
some new kinds of object, generically referred to as ``quantum black holes''
(see, e.g., Refs.~\cite{hsu,calmet}).
Since we do not have any experimental insight for such a high-energy regime,
it is conceptually difficult to conceive a theory for these objects, and it might
be a good starting point just to push our established knowledge to the limit.
\par
As we just recalled, it is of particular conceptual interest to study the possibility of black hole
production in high-energy collisions~\cite{eardleyG,Kanti:2008eq,Giddings:2001bu}.
Along these lines, Dvali and co-workers~\cite{dvali} recently went on to
conjecture that the high-energy limit of all physically relevant quantum field theories
involves the formation of a (semi)classical state (to wit, black hole formation
for gravity), which should automatically suppress trans-Planckian quantum fluctuations.
This idea extends the concept of a quantum uncertainty principle generalized
to include gravity, as was considered, for example in Refs.~\cite{scardigli,BNSminimallength}.
In this context, it can also be inferred that the mass of microscopic black holes must
be quantized, and admit a minimum value~\cite{dvalimin} (for more general cases,
see also Ref.~\cite{visser}).
\par
All of the above conjectures would be conspicuously substantiated if we could solve
the extremely complex dynamics of colliding Standard Model particles,
including the effect of the gravitational interaction, around the Planck scale~\cite{veneziano}.    
Since the completion of this task appears still far, we shall here study a toy model,
which assumes the validity of results known to hold in spherically symmetric
(classical) systems, and reproduces the above formula~\eqref{hoop} to a good approximation.
Bearing on this consistency check, we shall then be able to derive the existence of
a minimum black hole mass, provided there exists a minimum (fundamental or testable)
length~\cite{BNSminimallength,piero,Garay:1994en,Mead:1964zz,Calmet:2004mp}.
Beside a purely conceptual interest, the existence of this mass threshold
may have phenomenological implications in models with extra spatial
dimensions~\cite{add,rs}, where the fundamental (gravitational) length
corresponds to energy scales potentially as low as a few TeV's.
\section{Toy model for black hole production}
\setcounter{equation}{0}
Classical horizon formation is a relatively well-understood process in spherically
symmetric space-times.
We can write a general spherically symmetric metric $g_{\mu\nu}$ as
\be
\d s^2
=
g_{ij}\,\d x^i\,\d x^j
+
r^2(x^i)\left(\d\theta^2+\sin^2\theta\,\d\phi^2\right)
\ ,
\label{metric}
\ee
where $r$ is the areal coordinate and $x^i=(x^1,x^2)$ are coordinates
on surfaces with $\theta$ and $\phi$ constant.
The location of a trapping horizon is determined by the equation
(see, e.g.,~\cite{hayward} and references therein)
\be
0
=
g^{ij}\,\nabla_i r\,\nabla_j r
=
1-\frac{2\,m}{r}
\ ,
\label{th}
\ee
where $\nabla_i r$ is the covector perpendicular to surfaces of constant area
$\mathcal{A}=4\,\pi\,r^2$.
The quantity $m=\lp\,M/\mpl$ is the active gravitational (or Misner-Sharp)
mass function representing the total energy enclosed
within a sphere of radius $r$.
For example, is we set $x^1=t$ and $x^2=r$, the function $M$ is explicitly given
by the integral of the matter density $\rho=\rho(x^i)$
weighted by the flat metric volume measure,
\be
M(t,r)=\frac{4\,\pi}{3}\int_0^r \rho(t, \bar r)\,\bar r^2\,\d \bar r
\ ,
\label{M}
\ee
as if the space inside the sphere were flat.
Of course, it is in general very difficult to follow the dynamics of a given
matter distribution and verify the existence of surfaces satisfying Eq.~\eqref{th}.
\par
\begin{figure}[t]
\centering
\epsfxsize=9cm
\epsfbox{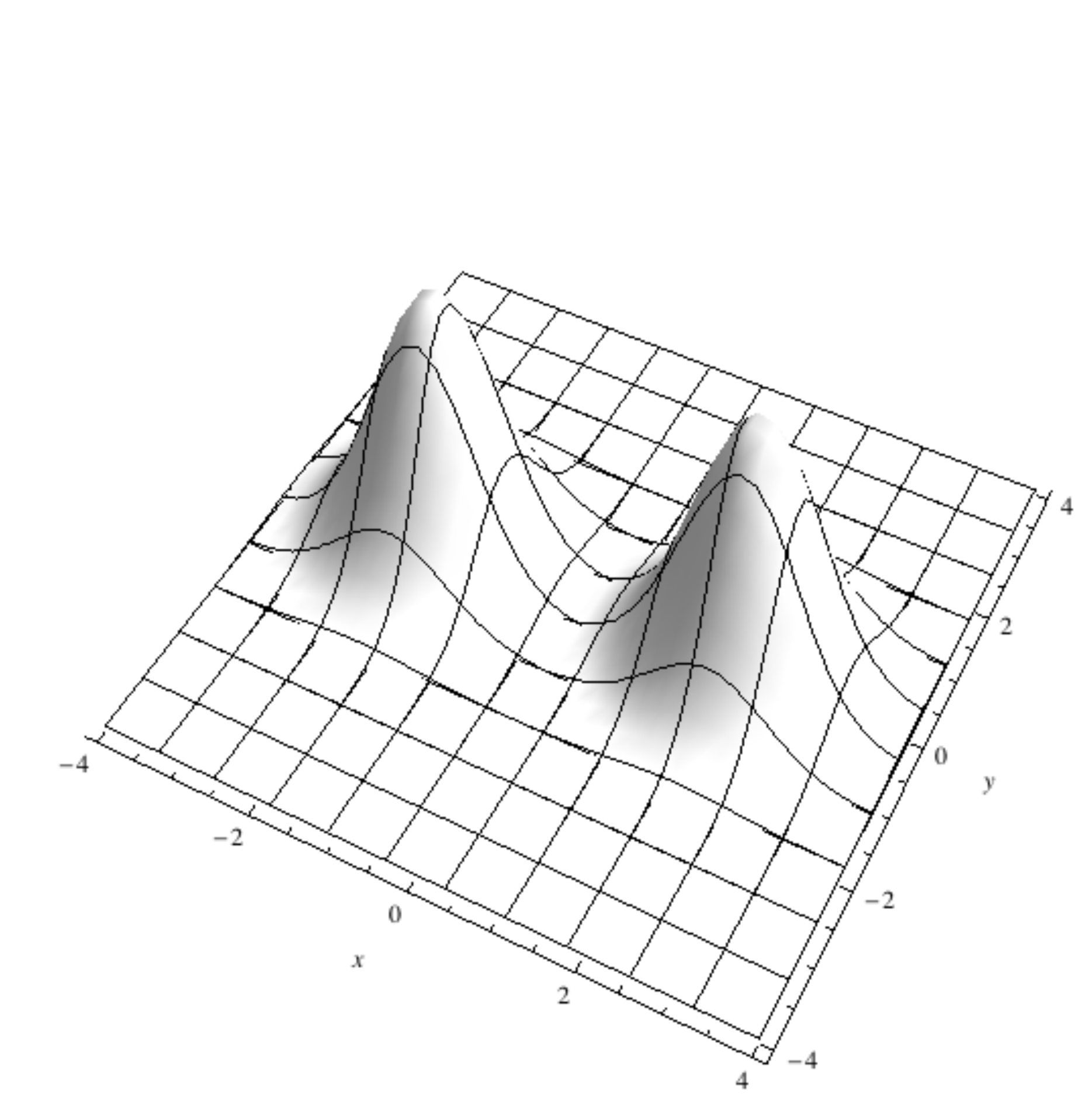}
\caption{Gaussian packets with $b=2\,\ell$ at their minimum distance.
All lengths are in units of $\ell$.
\label{gaussians}}
\end{figure}
Let us consider two equal Gaussian packets of width $\ell$ moving toward each other
in (asymptotically) flat space-time, say along the $z$ direction of a cartesian system
(see Fig.~\ref{gaussians}).
We neglect the details of their spatial extension along $z$, and just consider
the configuration of the two packets at the instant when their centers are
at the minimum distance, say $2\,b$.
We interpret the square of such packets as the ``energy density'' of two particles,
including the kinetic contribution,
\be
\rho_\pm(x,y)
&\!\!=\!\!&
\frac{\rho_0}{\pi\,\ell^2}\,
\exp\left\{-\frac{(x\pm b)^2+y^2}{\ell^2}\right\}
\nonumber
\\
&\!\!=\!\!&
\frac{\rho_0}{\pi\,\ell^2}\,
\exp\left\{-\frac{r^2\pm 2\,b\,r\,\cos(\theta)+b^2}{\ell^2}\right\}
=
\rho_\pm(r,\theta)
\ ,
\ee
where we introduced cylindrical coordinates on the plane transverse to $z$,
with origin at the midpoint between the packets, in $x=y=0$.
The Gaussians are normalized in the $x$-$y$ plane, so that the positive $\rho_0$
parametererizes the total energy, possibly encoding the details of the distribution
along the ``neglected'' direction $z$.
\par
If the system were spherically symmetric around the chosen origin, the metric
would reduce to the form~\eqref{metric}.
Its mass function~\eqref{M} would then be given by
\be
M(r)
&\!\!\propto\!\!&
\int_0^r
r'\,\d r'
\int_{-\pi}^{+\pi}
\d\theta
\left[\rho_+(r',\theta)+\rho_-(r',\theta)\right]
\nonumber
\\
&\!\!=\!\!&
4\,\pi\,\rho_0\,\int_0^r
\frac{r'\,\d r'}{\pi\,\ell^2}\,
e^{-\frac{{r'}^2+b^2}{\ell^2}}\,
I_0\!\left(\frac{2\,b\,r'}{\ell^2}\right)
\nonumber
\\
&\!\!=\!\!&
\rho_0\,\frac{\ell^2}{b^2}\,\int_0^{\frac{2\,b\,r}{\ell^2}}
s\,\d s\,
e^{-\frac{\ell^4\,s^2+4\,b^4}{4\,b^2\,\ell^2}}\,
I_0(s)
\ ,
\ee
where $I_0$ is the Bessel function of order $0$, and $\rho_0$ only
appears as a multiplicative factor with dimensions of a mass.
Let us assume we can still apply this concept of mass function
and proceed as if the system could be reasonably approximated
by arguments only properly defined in spherical symmetry.
\par
\begin{figure}[t]
\centering
\epsfxsize=8cm
\epsfbox{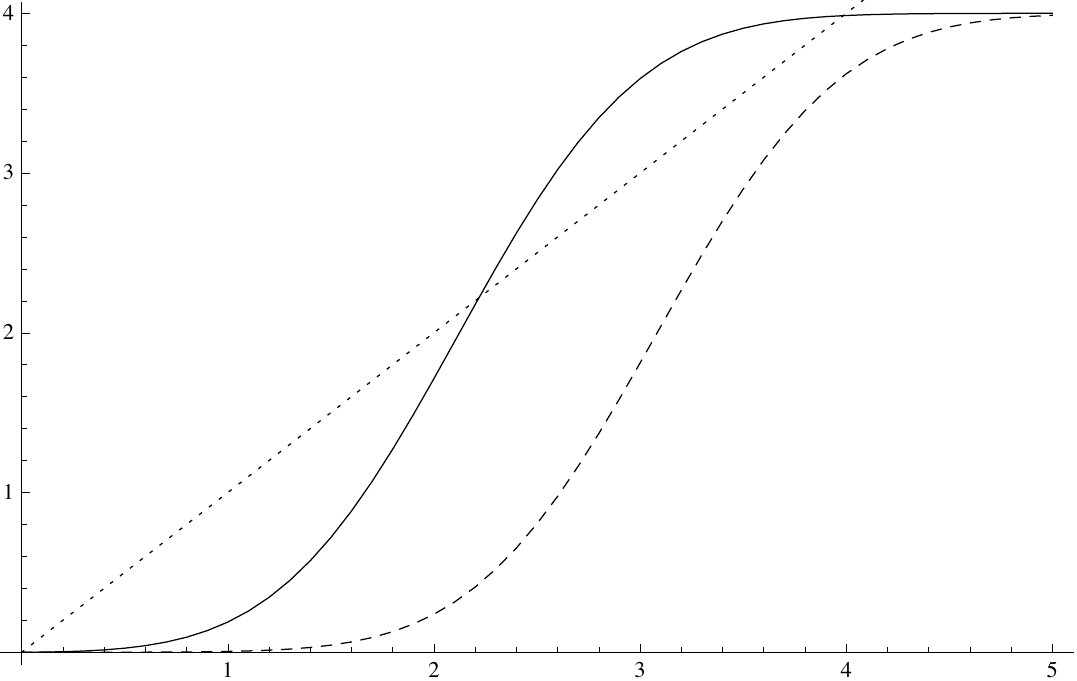}
\caption{Horizon function $R=R(r)$ for $\rho_0=\ell$
with $b=2\,\ell$ (solid line) and $b=3\,\ell$ (dashed line).
Possible horizons appear where the curves intersect the dotted line $R=r$.
All lengths are in units of $\ell$.
\label{BHnoBH}}
\end{figure}
The above integral can be performed numerically, for any given values of the parameters.
In Fig.~\ref{BHnoBH}, we plot the resulting horizon function
\be
R=\frac{2\,\lp\,M(r)}{\mpl}
\ ,
\ee
for $\rho_0=\ell$ with $b=2\,\ell$ (solid line) and $b=3\,\ell$ (dashed line).
The former case shows the appearance of a horizon where $R(r)=r$, whereas
the latter does not.
In fact, the horizon function $R=R(r)$ either intersects the axis $r$ twice or not at all.
The inner intersection, say $r=R_-$, occurs when  $R(r)<r$ grows faster than $r$,
whereas the second one, $r=R_+$, appears for $R(r)>r$ approaching the
asymptotic value.
Since $R(r)>r$ for $R_-<r<R_+$, it is the intersection $R_+$ which
might represent the outer horizon, and prevent signals originating from the system
from reaching out (whereas $R_-$ could be just a spurious artifact of our
toy model and we thus do not discuss it any further). 
Note that our calculation falls somewhat short of the hoop conjecture:
according to Eq.~\eqref{hoop}, $b=3\,\ell<4\,\ell$ should be close enough
for black hole formation, but this discrepancy could be easily removed by
suitably normalizing the Gaussian distributions so as to take into account the
spread along the $z$ direction more precisely.
\par
Upon varying the parameters $\ell$, $b$ and $\rho_0$, one comes to two
interesting results:
\begin{enumerate}
\item
For impact factors $b\lesssim \ell$, the horizon function quickly approaches an asymptotic
shape and is basically unaffected by the specific value of $b$ (see Fig.~\ref{blessell}).
Moreover, for $b\lesssim \ell$ the collision is almost head-on and the appearance of
a horizon seems to depend only on the ratio $\rho_0/\ell$;
\begin{figure}[ht]
\centering
\epsfxsize=8cm
\epsfbox{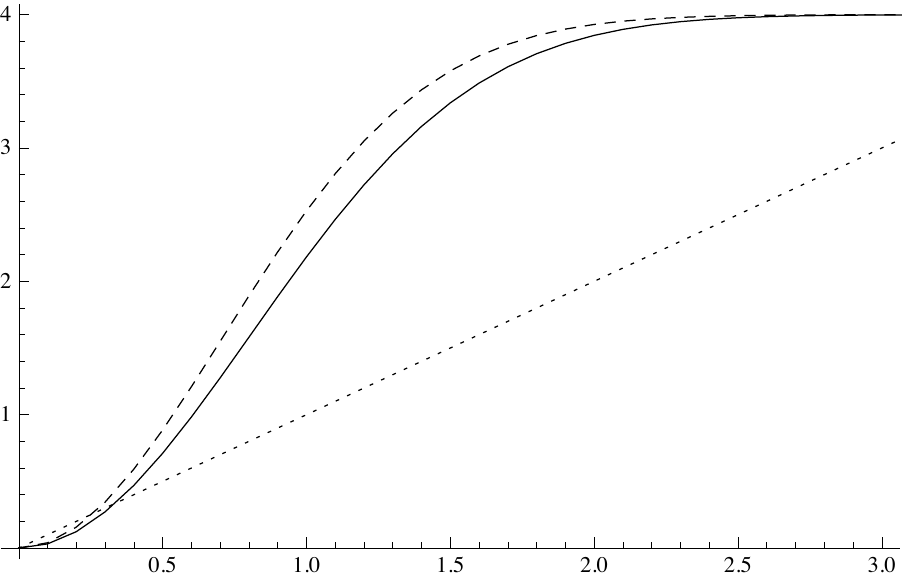}
\caption{Horizon function $R=R(r)$ for $\rho_0=\ell$
with $b=\ell/2$ (solid line) and $b=\ell/100$ (dashed line).
All lengths are in units of $\ell$.
\label{blessell}}
\end{figure}
\item
For fixed $\ell$ and $b$, there is a minimum value of $\rho_0$ (hence, a minimum
value of $M$, say $M_0$) which shows the presence of a horizon (see Fig.~\ref{minmass}).
\begin{figure}[ht]
\centering
\epsfxsize=8cm
\epsfbox{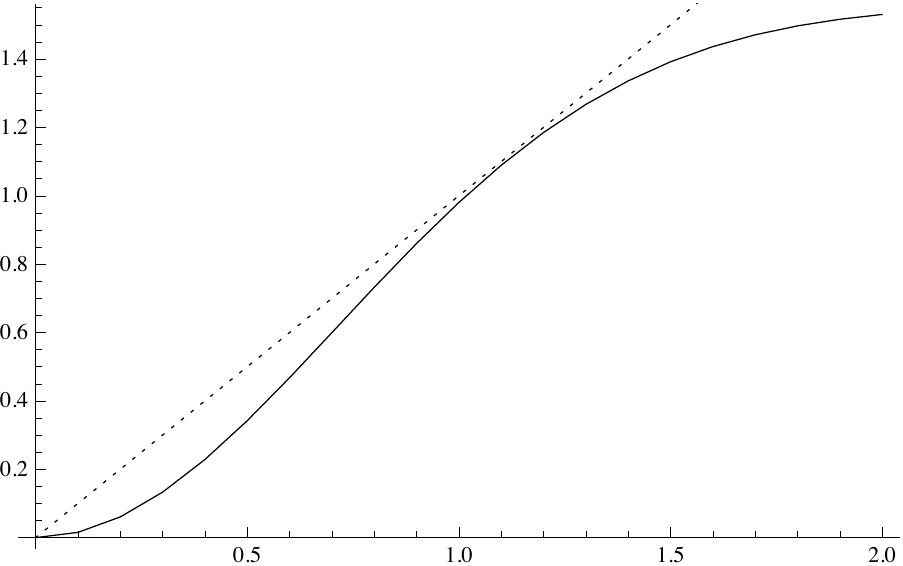}
\caption{Horizon function $R=R(r)$ for $b=\ell/10$ and $\rho\simeq 2\,\ell/5$.
All lengths are in units of $\ell$.
\label{minmass}}
\end{figure}
\end{enumerate}
Upon putting the two above results together, we can then draw the following
conclusion:
if there exists a minimum width $\ell$ for Gaussian densities, there is then a minimum
black hole mass
\be
M_0\simeq \mpl\left(\frac{\ell}{\lp}\right)
\ .
\label{m_0}
\ee
Of course, it would make little sense to infer a precise numerical factor,
given the simplicity of our calculation.
\par
In particular, effects of the gravitational interaction between the two packets 
on their evolution have not been accounted for.
This approximation might be still appropriate, if the packets do not significantly
deviate from the Minkowskian trajectories and tidal effects remain small.
Note, also, that we approximated the space-time geometry with a spherically
symmetric metric, which necessarily implies the Schwarzschild geometry sufficiently 
far away from the packets (say, for $r\gg \ell$).
However, for a nonzero impact parameter $b$, the resulting black hole must have
non-vanishing angular momentum (along the $y$ axis).
Moreover, the Gaussian packets may be meant to represent Standard Model particles
(or partons).
In our calculation, we did not assume any type of charges for the incoming waves
and did not have to deal with their conservation laws. 
A better description would generally require using the more complex Kerr(-Newman) metric.
Another point is that, if the energies of the two colliding packets are very different in the
chosen (inertial or freely falling) frame, the resulting object will have a mass smaller
than the sum of the energies of the two packets, part of the total energy
being recovered as kinetic energy~\cite{Calmet:2012mf}.
\par
Let us conclude this section by remarking that we have not assumed any specific
physical meaning for the width $\ell$.
The result~\eqref{m_0} is therefore general enough and can be applied to various
theories of space-time.
It is also clear that it will induce a modification in the black hole production cross
section, qualitatively similar to the low-energy suppression assumed in
Ref.~\cite{spallucci}.
\section{Discussion and physical implications}
\setcounter{equation}{0}
One can look at the simplified process of black hole production described in the previous
section from very different perspectives.
From a classical, general relativity point of view, one can think of the Gaussian functions
as the (smoothed) density profile of spatially extended systems,
such as galaxies or nebulae.
Clearly, if an amount of matter (dust, stars or energy in general) is concentrated inside
its horizon radius, a black hole will form and the rest of the system will be ``scattered''
according to the usual (but still very complex) general relativistic dynamics~\cite{payne}.
\par
However, if one is willing to extend the above model into quantum scales,
our view must change considerably.
In this context, one can either consider the Gaussian packets as ``probability distributions'',
as per point-particle quantum mechanics, or as describing the ``effective extension'' of 
elementary particles, as per quantum field theory.
In both cases, we must then ask ourselves what happens to the ``portion''
of particles that remains outside of the horizon.
Due to the indivisibility of such particles, there are two possibilities:
\par
\noindent
{\em i)\/} in the first case, one might think that ``if particles partly collapse, then they must collapse
entirely'' (perhaps with a probability proportional to the portion of wave-function included
within the horizon radius).
This means that the two Gaussian packets totally contribute to the formation of the black hole,
which will have a mass equal to the sum of the energies of the two colliding particles
in the chosen centre-of-mass frame, and a black hole is all that is left at the end of the process;
\par
\noindent
{\em ii)\/} in the second case, we can assume that only the part of the energy effectively included
within the horizon radius will be swallowed inside the black hole,
whereas the remaining energy is somehow scattered away in the shape of new particles.
As it can be seen from Fig.~\ref{BHnoBH}, the horizon function is larger than $r$
anywhere in the interval from $R_-$ to $R_+$.
It is thus in principle possible for a black hole with radius anywhere
in this range to form as a result of the collision, provided the difference between the
total center-of-mass energy of the incoming packets and black hole energy is scattered
in the form of other particles (maybe carrying away the Standard Model charges of
the incoming particles).
\par
\noindent
In both events, there is a mass threshold, but the spectrum of black hole masses
is continuous.
It has been suggested that black hole masses are instead quantized~\cite{dvalimin,visser}.
If this turned out to be the case, the second option seems more likely, since a black hole
would be created with quantized mass and respect energy conservation by scattering
away the remaining energy.
\par
In four space-time dimensions, it is plausible that $\ell\sim \lp$~\cite{scardigli}
and $M_0\sim\mpl\simeq 10^{16}\,$TeV.
This means that the above considerations remain rather speculative.
However, there are possible phenomenological implications in the context of models
in which gravity can also propagate along extra spatial dimensions.
If the gravitational scale $\mpl$ is within the reach of today's particle physics experiments,
as suggested by these scenarios, (quantum) black holes may be produced (and detected)
in colliders and by cosmic rays impinging on the atmosphere.
Our simple calculation then adds to the current picture by describing the collisions
between elementary particles, here viewed as objects extended in our three-dimensional
space, in the shape of Gaussian energy distributions.
If details of the extra-dimensional metric can be assumed to leave the picture
(qualitatively~\footnote{Of course, precise numerical coefficients will vary depending
on the bulk metric, but we shall leave a more detailed discussion of this issues for
future investigations.})
unaffected, we can again say that, if the impact parameter is small enough,
a black hole will form, confirming once more the hoop conjecture.
Moreover, what we stated above simply applies with a Planck scale lowered to
an energy potentially within a few TeV's.
In particular, it is possible that the entire initial energy will be transformed into
the black hole mass, but the resulting black hole can also have a mass below
this value.
In the latter case, the energy difference would be radiated in the form of Standard
Model particles or bulk gravitons. 
The phenomenological implications for black holes which are created and decay
instantaneously are then not clear.
If, however, black holes live (relatively) long~\cite{Casadio:2009sz,atlasBO},
this result implies two possible signatures:
\par
\noindent
{\em a)\/} on the one hand, when black holes are formed with masses near the maximum
possible value (their horizon radius being approximately equal to $R_+$),
the entire available energy falls into the black hole and no other particles are radiated.
In this case the resulting black holes must carry all the Standard Model charges of
the incoming particles.
Their signatures in particle accelerators would thus be large quantities of
missing energy and transverse momentum, associated with the simultaneous
detection of a very heavy and charged particle.
This would be a very distinctive event, if we assume that the incoming Gaussian
packets represent two quarks, since the resulting electric charge will have a fractional
value~\cite{moedal};
\par
\noindent
{\em b)\/} on the other hand, if the black holes were formed with masses bellow the maximum
value, the charge could be radiated in the form of Standard Model particles,
and the only sign of the black hole existence would just be large amounts of missing
energy and transverse momentum.
\par
Let us conclude by remarking again that the angular momentum and charges of the colliding
particles are totally missing in the final black hole state we used for our derivation.
It is possible that the angular momentum is emitted during the collision, like the initial
Standard Model charges.
However, if the black hole retains (part of) it, it will be very important to determine
its effects on the mass threshold~\eqref{m_0}, along with those induced by any gauge
charges.
These points are left for future investigations.
\section*{Acknowledgements}
This collaboration was made possible by the COST--Action~MP0905.
O.M.~was supported by UEFISCDI research grant PN-II-RU-TE-2011-3-0184.
\end{document}